\documentclass[useAMS,usenatbib]{mn2e}

\usepackage[psamsfonts]{amssymb}
\usepackage[dvips]{graphicx}
\usepackage{amsmath,alltt}                                                                                                                          
\usepackage{multirow}
\usepackage{rotating}
\usepackage{lscape}       
\title[Back to the Future]{Back to the Future: Estimating Initial Globular Cluster Masses from their Present Day Stellar Mass Functions}
\author[Webb $\&$ Leigh]{Jeremy J. Webb$^{1}$, Nathan W. C. Leigh$^{2}$$^,$$^{3}$
\thanks{E-mail: webbjj@mcmaster.ca (JW), nleigh@amnh.org (NL)} \\
$^{1}$McMaster University, Department of Physics and Astronomy, 1280 Main St. W., Hamilton, Ontario, Canada, L8S 4M1 \\
$^{2}$Department of Astrophysics, American Museum of Natural History, Central Park West and 79th Street, New York, NY 10024 \\
$^{3}$Department of Physics, University of Alberta, CCIS 4-183, Edmonton, AB T6G 2E1, Canada}

\begin{document}

\pagerange{\pageref{firstpage}--\pageref{lastpage}} \pubyear{2015}

\maketitle

\label{firstpage}

\begin{abstract}

We use $N$-body simulations to model the 12 Gyr evolution of a suite of star clusters with identical initial stellar mass functions over a range of initial cluster masses, sizes, and orbits. Our models reproduce the distribution of present-day global stellar mass functions that is observed in the Milky Way globular cluster population. We find that the slope of a star cluster's stellar mass function is strongly correlated with the fraction of mass that the cluster has lost, independent of the cluster's initial mass, and nearly independent of its orbit and initial size. Thus, the mass function - initial mass relation can be used to determine a Galactic cluster's initial total stellar mass, if the initial stellar mass function is known. We apply the mass function - initial mass relation presented here to determine the initial stellar masses of 33 Galactic globular clusters, assuming an universal Kroupa initial mass function. Our study suggests that globular clusters had initial masses that were on average a factor of 4.5 times larger than their present day mass, with three clusters showing evidence for being 10 times more massive at birth.

\end{abstract}


\section{Introduction} \label{intro}

Globular clusters are old, ($\sim 10-12$ Gyrs \citep{marinfranch09}) spherical, gravitationally-bound configurations of stars with total present-day masses in the range $\sim$ 10$^4$-10$^6$ M$_{\odot}$. These clusters are believed to have formed within the high pressure environments of the primordial Milky Way, and mark some of the initial sites of star formation \citep{kruijssen14}. Having been found in all types of galaxies, we know they must play a pivotal role in the galaxy formation process in general. Once a cluster forms, stellar evolution and dynamics result in many of the cluster's initial properties changing over time \citep[e.g.]{gnedin99,fall01,gieles08,webb14a,brockamp14}. Star cluster simulations allow us to trace the evolution of clusters over their entire lifetime, and let us compare their present day properties to their primordial state. Understanding the initial conditions under which a globular cluster population forms allows us to explore how galaxies form and further constrain galaxy formation models.

Studies regarding the dynamical evolution of a globular cluster in isolation date back to \cite{henon61}. After an initial phase of mass loss via stellar evolution, the dynamical evolution of star clusters is driven by two-body relaxation \citep[e.g.][] {henon61, henon73, spitzer87, heggie03, gieles11}. Relaxation drives mass segregation, causing high-mass stars to sink deeper in the cluster potential and low-mass stars to drift outward. In terms of structure, the cluster will slowly expand and dissolve, however the timescale for dissolution in the absence of a tidal field is much greater than a Hubble time. The presence of a tidal field accelerates the dissolution time of a cluster, as two-body interactions only need to push stars beyond the tidal boundary as opposed to speeds greater than the cluster's escape velocity. The tidal boundary will fluctuate over the course of a cluster's orbit if it is non-circular or inclined. 

In a realistic gravitational potential, clusters will also be subject to tidal shocks which inject energy into the cluster and temporarily minimize the tidal boundary during perigalactic passes, passages through spiral arms and the Galactic disk, and encounters with giant molecular clouds (GMCs). In fact, \citet{lamers06} finds that encounters with GMCs is the dominant mechanism behind evaporation for clusters orbiting in the solar neighborhood. Simulations by \citet{kruijssen11} of clusters with and without tidal shocks illustrate that tidal shocks will lead to between $80-85\%$ additional dissolution events depending on the galaxy model. While tidal shocks experienced by clusters within the high-pressure environment in which they form would lead to very short dissolution times, the distribution of clusters in the Milky Way indicate that clusters survive by migrating outwards into the halo. The migration can likely be attributed to galaxies forming and growing via the hierarchical merger of smaller galaxies \citep{springel05}, with merger events pushing clusters out to wider orbits \citep{kruijssen14}. However as \citet{kruijssen12} points out, tidal shocks via merger events also have the ability to significantly disrupt clusters, sometimes to the point of complete dissolution. If a cluster survives to reach its present day orbit in the halo, the tidal shocks experienced by the cluster will be much weaker than when it was in its formation environment, with the dissolution times of surviving clusters becoming greater than a Hubble time.

When a cluster is stripped of stars, regardless of whether the mass loss event is due to two-body interactions, tidal stripping, or tidal shocks, it will primarily be stars orbiting in the outskirts that are removed. Hence over time, since two-body relaxation leads to mass segregation, the probability that a given star will escape from a cluster into the tidal field of its host galaxy over the course of a relaxation time increases as a function decreasing mass. In turn, the mean stellar mass in clusters slowly increases over time, and with a rate that decreases with increasing total cluster mass since the time-scale for two-body relaxation scales as N$^{1/2}$, where N is the number of objects in the cluster.   

\citet{vesperini97} conducted one of the first studies that explored the effects of different evolutionary processes on a cluster's stellar mass function. The authors considered the effects of relaxation, stellar evolution, disc shocking and a tidal field on globular cluster evolution. A key result from \citet{vesperini97} is that the difference between the initial and present day mass function of a cluster is primarily dependent on the fraction of mass lost, with only minor scatter due to initial mass, initial concentration, and Galactocentric distance (assuming circular orbits). This finding was confirmed by \citet{kruijssen09b} and \citet{trenti10}, however the former noted that the retention fraction of black holes could alter the evolution of the mass function. More specifically, a cluster hosting either a central massive black hole or multiple stellar-mass black holes will have a higher escape rate of high-mass stars due to their interaction with black holes in the cluster core. The mass function will in turn evolve at a much slower rate as low-mass stars are no longer the only stars escaping the cluster \citep{lutzgendorf13}. \citet{kruijssen09b} also finds that the evolution of the mass function can be dependent on the dissolution time of the cluster, as the mass function of young clusters with short dissolution times evolves slowly if most of the mass loss is occurring before high mass stars reach the end of their life.

\citet{webb14a} builds on previous work and explored the effects of orbital eccentricity on the evolution of the stellar mass function slope $\alpha$. The main result of \citet{webb14a} is that for a given perigalactic distance $R_p$, the dynamical evolution of the mass function slows with increasing orbital eccentricity due to the weaker mean tidal field it experiences. Thus the slope of the mass function $\alpha$ increases (i.e. becomes less negative) over time, with a rate that increases as the mean strength of the tidal field increases. When comparing model clusters with the same fraction of initial mass lost, the evolution in $\alpha$ was almost orbitally-independent. For a given fraction of mass lost there was only a minor spread in $\alpha$ due to cluster orbit as stronger tidal fields can remove higher mass stars from the cluster (i.e. the dependence of the probability of escape on stellar mass becomes slightly weaker in a stronger tidal field). This result further strengthens the possible relationship between the mass function of a globular cluster at a given time and the fraction of its initial mass that has been lost. However, we noted in \citet{webb14a} that differences due to the combined effects of orbital eccentricity, orbital inclination, initial size, and initial mass of a globular cluster may in some cases slightly alter the relation. 

The present paper is aimed towards quantifying how these combined effects will alter the evolution of $\alpha$ as a function of fraction of initial mass lost. \citet{lamers13} notes that comparing clusters as a function of initial mass lost as opposed to time ensures that clusters are being compared on the same evolutionary timescale. If a globular cluster's stellar mass function can be shown to be primarily dependent on the fraction of mass lost, it will also represent the clusters's dynamical age. Hence the initial mass of a cluster can be determined assuming that all globular clusters formed with the same stellar initial mass function (IMF) and either:
 
 \begin{itemize}
  \item A) tidal shocks experienced by clusters before they migrate to their present day orbit remove stars over the entire mass spectrum, such that the mass function does not evolve. Hence all clusters will still have the same IMF once they reach the halo and their mass function will reflect their dynamical age since escaping from their formation environment. If this is the case, the initial cluster mass which corresponds to the cluster's present day mass function will represent its mass at the time of migration into the halo (i.e. its current orbit). Or, 
 
 \item B) tidal shocks experienced by clusters before they migrate to their present day orbit remove stars in a similar manner as tidal evaporation and disk shocking, just over a shorter time scale. Hence the evolution of the mass function will remain unchanged with respect to the fraction of mass lost, and the mass function will reflect a cluster's dynamical age since formation. In this scenario, the initial cluster mass calculated from the cluster's present day mass function will represent its true initial mass at formation.

 \end {itemize}
 
\noindent However if tidal shocks that occur while clusters are still within their formation environment remove stars such that the evolution of the mass function is affected in a manner different from the above picture, then the connection between a cluster's stellar mass function and its dynamical age weakens. We will henceforth refer to this kind of mass loss as "non-standard", since the underlying physical mechanisms are unknown.
 
The assumption that all clusters form with the same IMF is consistent with the observed present-day stellar mass functions, once their dynamical evolution has been accounted for \citep{leigh12}. Hence the possibility of simply using the mass function of present day globular clusters to predict their initial masses becomes an exciting possibility. How exactly globular cluster populations evolve from their initial cluster mass function (GCIMF) to the present-day observed globular cluster mass function (GCMF) is still a topic of much debate.

The GCMF has been found to be nearly universal, a gaussian centred around a mean mass of approximately $10^5 M_\odot$ \citep{brodie06}. Whatever its initial form, the GCIMF will have undergone significant evolution due to clusters being subjected to tidal shocks within the formation environment \citep{kruijssen14}, internal two-body relaxation, dynamical friction, disk shocks and bulge shocks \citep{gnedin99}. It is not clear whether the present day mass function contains any traces of the GCIMF. The GCIMF is often assumed to have a power-law form, as this would would match the mass functions of other astronomical objects like molecular clouds and young clusters \citep{fall01}. A similar conclusion was reached by \citet{mclaughlin08}, who finds that the Galactic GCMF depends on cluster half-mass density, which they argue is a signature of a mass function that initially rose towards low masses but has eroded over time due to internal two-body relaxation driving the dissolution of its (preferentially low-mass) star clusters. However both these early studies make the simplifying assumption that cluster mass-loss rates are environmentally-independent. \citet{gieles08} argued that a power law GCIMF will not evolve to become the present-day GCMF due to internal two-body relaxation alone, whereas other analytic prescriptions and globular cluster models have been able to reproduce the present-day GCMF by including not only the effects of evaporation by two-body relaxation, but also gravitational shocks due to a tidal field, stellar evolution, dynamical friction, gas expulsion and radial anisotropy \citep{fall01,vesperini01,brockamp14}. Other studies have also been able to reproduce the present-day GCMF beginning with a log-normal GCIMF \citep{vesperini98}. All of these studies are based on treating the present-day GCMF and the GCIMF as probability distribution functions which can be altered due to several processes. A method for determining the initial mass of the remaining globular clusters in the Milky Way would provide robust constraints on the possible formation of globular cluster systems and galaxies themselves, while also providing insight into whether we understand how the various mechanisms discussed above alter the GCIMF over time.

In this study, we combine the results presented in \citet{webb14a} with additional models that range in initial mass, initial size and orbit in order to develop a method of predicting the initial total stellar mass of a globular cluster based on its present-day stellar mass function. The effects of two-body relaxation are quantified by considering a range of cluster masses and sizes. By placing our model clusters on a range of orbits we also consider the effects of tidal heating, tidal shocks, and the initial degree of tidal-filling. We also allow stars in our model cluster to evolve, hence stellar evolution is also being considered. In Section \ref{nbody} we introduce the models used in this study and in Section \ref{mfim} we compare the value of $\alpha$ for each model cluster at different time steps to the fraction of mass each cluster has lost. In Section \ref{discussion} we will quantify the relationship between $\alpha$ and the fraction of mass lost by a cluster and use it to predict the initial masses of select Galactic globular clusters. We summarize our findings in Section \ref{summary}.

\section{N-body models} \label{nbody}

The direct $N$-body code NBODY6 \citep{aarseth74,aarseth03} is used to model star clusters over 12 Gyr of evolution. We use a Plummer density profile \citep{plummer11,aarseth74} to generate the three dimensional position and velocity of each star within the cluster out to a cut-off of $\sim 10 \ r_m$, where $r_m$ is the system half-mass radius. Since we are primarily concerned with the evolution of the stellar mass function, the stars within each model cluster are assigned masses based on a \citet{kroupa93} IMF (i.e. the IMF found by \citet{leigh12} that best reproduces the observed cluster-to-cluster variations in the present-day stellar mass functions in a large sample of Galactic GCs) between 0.1 and 30 $M_{\odot}$. The \citet{kroupa93} IMF has the functional form:

\begin{equation}\label{eqn:mfunc}
\frac{dN}{dm} = m^\alpha
\end{equation}

where $\alpha$ equals -2.7 for $m > 1 M_\odot$, -2.2 for $0.5 \le m \le 1  M_\odot$, and -1.3 for $0.08 < m \le 0.5  M_\odot$. With respect to Equation \ref{eqn:mfunc}, the commonly cited Salpeter IMF has an $\alpha$ of -2.35 \citep{salpeter55}.

Model clusters all have metallicities of $Z=0.001$ and binary fractions of $4\%$. Binary stars are assigned total masses based on two stars drawn from the IMF, while the mass ratio between the two stars are selected from a uniform distribution. The initial distribution of binary periods matches that of \citet{duquennoy91} while the initial distribution of orbital eccentricities follow a thermal distribution \citep{heggie75}. Details regarding the stellar and binary evolution algorithms used in our simulations can be found in \citet{hurley08a} and \citet{hurley08b}. In order to compare the results of our models to Galactic globular clusters, our model clusters orbit within a Milky Way-like potential comprised of a $1.5 \times 10^{10} M_{\odot}$ point-mass bulge, a $5 \times 10^{10} M_{\odot}$ \citet{miyamoto75} disk (with $a=4.5\,$kpc and $b=0.5\,$kpc), and a logarithmic halo potential \citep{xue08}. The logarithmic halo potential is scaled such that the circular velocity at a galactocentric distance of $8.5\,$kpc is 220 km/s. The majority of models discussed in this study were initially presented in either \citet{webb13}, \citet{leigh13}, or \citet{webb14b}. Additional models only differ in initial mass and/or size. We refer the reader to these previous studies for a more detailed description of our simulations.

\begin{figure}
\centering
\includegraphics[width=\columnwidth]{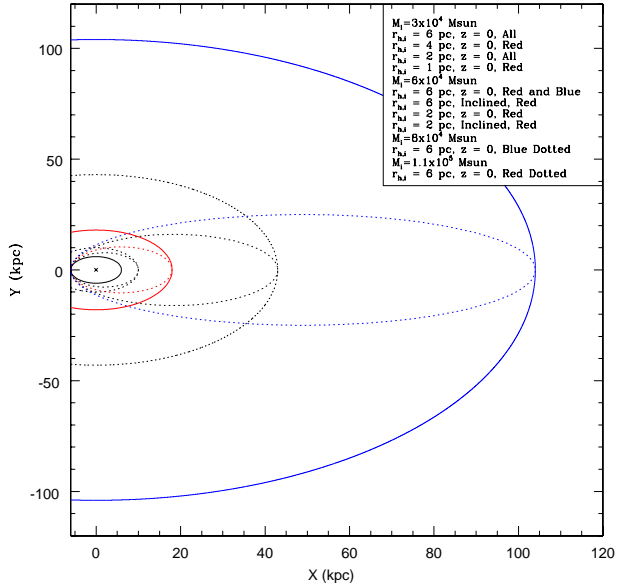}
\caption{Illustration of star cluster orbits that are used in this study. Each initial mass and size combination was modelled with a circular orbit at 6 kpc (solid black line). Additional orbits that correspond to various initial mass and size combinations are marked in the legend. For inclined orbits, star clusters were modelled with orbital inclinations of $22^\circ$ and $44^\circ$.}
  \label{fig:orbits}
\end{figure}

Our initial suite of simulated clusters had initial masses of either $3 \times 10^4 M_\odot$ or $6 \times 10^4 M_\odot$ and initial half-light radii $r_h$ of either 2 pc or 6 pc. The models are taken from previous works studying the effects of initial size and orbit on the dynamical evolution of star clusters \citep{webb13, leigh13, webb14a}. To further study the effects of initial mass on cluster evolution, select clusters were re-simulated but with initial masses of $8 \times 10^4 M_\odot$ and  $1.1 \times 10^5 M_\odot$. For a given initial mass and size, we model clusters with orbital eccentricities of 0, 0.25, 0.5, 0.75, and 0.9, where eccentricity is defined as $e = \frac{R_{a}-R_p}{R_{a}+R_p}$. $R_{a}$ and $R_{p}$ are the apogalactic and perigalactic distances of the orbit, respectively. All models with eccentric orbits have a $R_p$ of 6 kpc. For comparison purposes, a cluster was also modelled with a circular orbit at the $R_a$ of each eccentric cluster.  From a previous study focusing on the effects of orbital inclination on cluster evolution \citep{webb14b}, we also include model clusters with orbits having inclinations of $22^\circ$ and $44^\circ$ relative to the plane of the Galaxy. These additional models also include clusters with inclined \textit{and} eccentric orbits. Since our previous model clusters with inclined orbits all had an initial $r_h$ of 6 pc, we have re-simulated these clusters with an initial $r_h$ of 2 pc to explore the effects of initial size on cluster evolution. The initial model parameters of our entire suite of simulations are summarized in Table \ref{table:modparam}. The orbits covered by our suite of simulations are illustrated in Figure \ref{fig:orbits}. \footnote{It should be noted that all model clusters have orbital velocities greater than 88 mph.}

\begin{table}
  \caption{Model Input Parameters}
  \label{table:modparam}
  \begin{center}
    \begin{tabular}{lcccccc}
      \hline\hline
      {$M_i$} &{$r_{h,i}$} & {$R_p$} & {e} & {i} & {Reference} \\
      {$M_\odot$} & {pc} & {kpc} & { } & {degrees} & {} \\
      \hline

$3 \times 10^4$ & 2 & 6 & 0 & 0 & \citet{webb13} \\
{} & {} & 6 & 0.25 & 0 & {}\\
{} & {} & 10 & 0 & 0 & {}\\
{} & {} & 6 & 0.5 & 0 & {}\\
{} & {} & 18 & 0 & 0 & {}\\
{} & {} & 6 & 0.75 & 0& {}\\
{} & {} & 43 & 0 & 0& {}\\
{} & {} & 6 & 0.9 & 0 & {}\\
{}& {} & 104 & 0 & 0 & {}\\

$3 \times 10^4$ & 6 & 6 &  0 & 0 & \citet{webb13} \\
{} & {} & 6 & 0.25 & 0 & {}\\
{} & {} & 10 & 0 & 0 & {}\\
{} & {} & 6 & 0.5 & 0& {}\\
{} & {} & 18 & 0 & 0& {}\\
{} & {} & 6 & 0.75 & 0 & {}\\
{} & {} & 43 & 0 & 0 & {}\\
{} & {} & 6 & 0.9 & 0& {}\\
{}& {} & 104 & 0 & 0 & {}\\

$6 \times 10^4$ & 6 & 6 & 0 & 0 &\citet{leigh13} \\
{} & {} & 6 & 0.5 & 0 & {}\\
{} & {} & 18 & 0 & 0 & {}\\
{} & {} & 6 & 0.9 & 0 & {}\\
{}& {} & 104 & 0 & 0 & {}\\

{} & {2} & 6 & 0 & 22 & This publication\\
{} & {} & 6 & 0.5 & 22& {} \\
{} & {} & 18 & 0 & 22& {} \\
{} & {} & 6 & 0.9 & 22& {} \\
{}& {} & 104 & 0 & 22& {} \\
{} & {} & 6 & 0 & 44 & {} \\
{} & {} & 6 & 0.5 & 44& {} \\
{} & {} & 18 & 0 & 44& {} \\
{} & {} & 6 & 0.9 & 44& {} \\
{}& {} & 104 & 0 & 44& {} \\

{} & {6} & 6 & 0 & 22 &\citet{webb14b} \\
{} & {} & 6 & 0.5 & 22& {} \\
{} & {} & 18 & 0 & 22& {} \\
{} & {} & 6 & 0.9 & 22& {} \\
{}& {} & 104 & 0 & 22& {} \\
{} & {} & 6 & 0 & 44 & {} \\
{} & {} & 6 & 0.5 & 44& {} \\
{} & {} & 18 & 0 & 44& {} \\
{} & {} & 6 & 0.9 & 44& {} \\
{}& {} & 104 & 0 & 44& {} \\

$8 \times 10^4$ & 6 & 6 &  0 & 0 & This publication \\
{} & {} & 6 & 0.9 & 0 & {}\\

$1.1 \times 10^5$ & 6 & 6 &  0 & 0 & This publication \\
{} & {} & 6 & 0.5 & 0 & {}\\


      \hline\hline
    \end{tabular}
  \end{center}
\end{table}

\section{The Stellar Mass Function - Initial Mass Relation}\label{mfim}

To find the value of $\alpha$ in Equation \ref{eqn:mfunc} for each model cluster at a given time step, we first mark all main sequence stars within the tidal radius of the cluster that have masses between $0.15 M_\odot$ and $0.8 M_\odot$. The latter mass corresponds approximately to the main sequence turn off at 12 Gyr. Excluding main sequence stars beyond the main sequence turn off has a negligible effect on the slope of the mass function, since these phases of stellar evolution represent a very narrow range in stellar mass. This exclusion also ensures that we are not including main sequence stars that have formed via collisions or mergers. In order for our models to be compared to observations, we assume binary stars are unresolved and treat them as single stars. The total mass of the binary is calculated from the total luminosity and assuming $L \sim M^{\frac{1}{3}}$. We then set $\alpha$ equal to the slope of a plot of$log(\frac{dN}{dn})$ versus $log(m)$. It should be noted that since the mass function is best represented by a two-part power law, we take $\alpha$ to be the slope of the high mass end of the mass function, beyond $0.15 M_\odot$.

\begin{figure}
\centering
\includegraphics[width=\columnwidth]{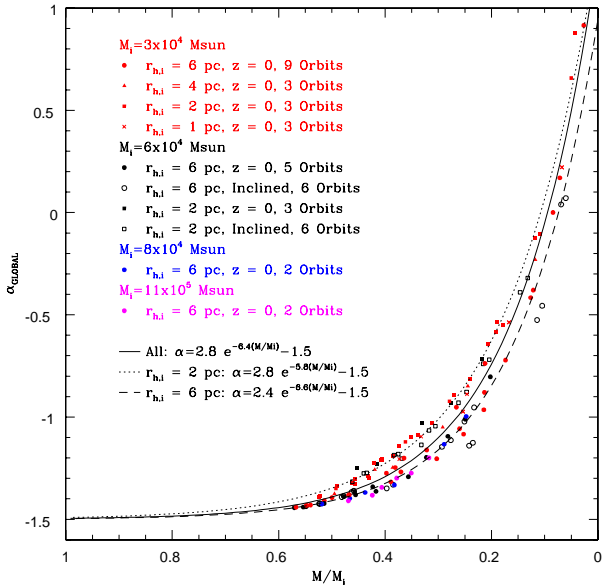}
\caption{Slope of the stellar mass function at 6, 10, and 12 Gyr versus the fraction of mass lost for each model cluster. Points are colour coded based on initial mass while different symbols represent different initial sizes as indicated. Exponential curves of best fit are matched to all data points (solid line), clusters with initial sizes of 6 pc (dotted line) and 2 pc (dashed line).}
  \label{fig:alpha}
\end{figure}

To get a sense of how the mass function of each model cluster evolves, we plot $\alpha$ for each model at 6, 10, and 12 Gyr as a function of fraction of mass lost in Figure \ref{fig:alpha}. It should be noted that some of the lower initial mass models reach dissolution before 12 Gyr, and therefore only have two data points in the figure.

As expected, the general trend is for $\alpha$ to increase as a cluster loses mass due to either two-body interactions ejecting stars from the cluster or tidal stripping and bulge and disc shocks removing primarily low mass stars that have been mass segregated to the outer regions \citep[e.g.][]{leigh12}. In order to track the evolution of $\alpha$ as a function of mass lost, we fit our data points with an exponential curve of the form:

\begin{equation}\label{mfrelation}
\alpha = A \times e^{B \frac{M}{M_i}}+C
\end{equation}

In the absence of stellar evolution and binary stars, the offset C would simply be the initial slope of the mass function $\alpha_i$. Since we include the effects of stellar evolution and binaries here, the offset C must compensate for $\alpha$ initially decreasing due to the cluster losing mass via stellar evolution and the disruption of binary stars \citep{webb14a}. During this phase, which lasts until clusters have lost approximately $40\%$ of their initial mass, clusters will also expand and lose some mass in the form of stars. For our models, C is set to be just larger than the most negative value of $\alpha$. Since we do not model the initial decrease in $\alpha$, our relation is only applicable to star clusters that have lost greater than $40\%$ of their initial mass. This criterion is met by Galactic globular clusters which will have lost a significant amount of their initial mass from stellar evolution alone.

Initially taking into consideration all models, we find that $A=-2.7 \pm 0.1 $ and $B = -6.4 \pm 0.1$, which is plotted as a solid curve in Figure \ref{fig:alpha}. Here we have established an evolutionary track for $\alpha$ that is solely dependent on the amount of mass a cluster has lost. However, \citet{kruijssen09b} notes that the evolution in $\alpha$ will be different for clusters with short dissolution times. Comparisons of Equation \ref{mfrelation} to the work of \citet{lutzgendorf13} suggests that the existence of a massive black hole (greater than $1\%$ of the total cluster mass) at the centre of a cluster may alter our prediction of $\frac{M}{M_i}$ by up to $20 \%$. Observations of such massive black holes in globular clusters have yet to be confirmed. \citet{lutzgendorf13} also finds that the black hole retention fraction can alter the evolution of $\alpha$, however the added uncertainty in $\frac{M}{M_i}$ of $\sim 10\%$ is comparable to the scatter about Equation \ref{mfrelation}, as discussed below.

The scatter about the relation given by Equation \ref{mfrelation} is only $\sigma=0.09$, and is primarily due to cluster orbit and initial cluster size, while initial cluster mass appears to have no affect on the evolution of $\alpha$. We recover the actual initial mass of each of our models to within $7 \%$ when using Equation \ref{mfrelation}. As previously found in \citet{webb14a}, cluster orbit can slightly alter the evolution of $\alpha$ since stronger tidal fields can remove higher mass stars than weaker tidal fields. When considering clusters of different initial size, a larger amount of scatter is introduced to the mass function - initial mass relation. An under-filling cluster will evolve as if it is in isolation, undergoing a certain degree of two-body relaxation and mass segregation until it expands to the point of becoming tidally affected \citep{gieles11, alexander13, webb13}. This is qualitatively equivalent to assuming some degree of primordial mass segregation in a tidally-filling cluster. When an initially under-filling cluster begins losing stars it is preferentially losing low mass stars that have been segregated to the outer regions, which is reflected in a faster increase in $\alpha$ as a function of the fraction of mass lost. Since a tidally filling cluster is subject to tidal stripping almost immediately, a mix of high and low mass stars is removed before segregation becomes important, resulting in a shallower increase in $\alpha$ initially. 

If we choose to fit the extended (circles - $r_{h,i} = 6$ pc) and compact (squares - $r_{h,i} = 2$ pc) clusters separately, we find that for extended clusters $A=-2.4 \pm 0.1 $ and $B = -6.6 \pm 0.1$ (plotted as a dashed line). With the exception of the clusters orbiting beyond 18 kpc, all extended models are tidally filling at $R_p$ and for a higher fraction of their orbit if their eccentricity is low. Hence these clusters can lose stars over a higher mass range initially, slowing the evolution of $\alpha$ compared to the overall population. For compact clusters we find $A=-2.8 \pm 0.1 $ and $B = -5.8 \pm 0.1$ (plotted as a dotted line). All of the compact clusters are initially tidally under-filling, such that they have time to undergo some degree of mass segregation before tidal stripping becomes a major source of mass loss. Hence the evolution of $\alpha$ is faster than the overall population since primarily low-mass stars are being removed. It should be noted that the uncertainty in the extended and compact best fit relations are $5\%$ and $2.5\%$ respectively, indicating that knowing a cluster's initial size will only marginally improve the determination of its initial mass via the mass function - initial mass relation presented here. However as we will discuss in Section \ref{discussion}, the uncertainty due to the omission of additional mass loss mechanisms from our models is expected to dominate compared to the uncertainty in initial cluster size.

We draw the reader's attention to the specific set of extended data points that are significantly removed from the dashed line of best fit in Figure \ref{fig:alpha}. These open circles represent tidally-filling clusters with circular orbits at 6 kpc but with orbital inclinations of $22^\circ$ and $44^\circ$. As found in \citet{webb14b}, clusters with inclined orbits at low $R_{gc}$ are subject to significant amounts of tidal heating and disk shocking which result in significant stellar mass loss. Due to the strength of the tidal field and the fact that these clusters are not initially mass segregated, higher mass stars can be removed from the cluster than if its orbit was not inclined, slowing the evolution of $\alpha$. Since these clusters deviate slightly from our derived mass function - initial mass relation, we suggest that if the orbit of a cluster is known to be inclined and at low Galactocentric radius, the line of best fit found by considering only our initially extended models will yield the initial cluster mass to within a few percent.

The best fit values to both the entire dataset as well as the extended and compact clusters are listed in Table \ref{table:alphaparam}. In order to apply our relation to a wider range of observational datasets, we also include in Table \ref{table:alphaparam} the best fits values of A, B, and C when only stars within $0.1 - 0.5 M_\odot$, $0.3 - 0.8 M_\odot$, or  $0.5 - 0.8 M_\odot$ are used to find $\alpha$. Using the lower mass range our relation recovers the initial mass of each model cluster to within $8.6\%$ and using the high mass range recovers the initial mass of each model cluster to within $7\%$.

\begin{table}
  \caption{Model Input Parameters}
  \label{table:alphaparam}
  \begin{center}
    \begin{tabular}{lcccc}
      \hline\hline
      {$Dataset$} &{$A$} & {$B$} & {C} \\
      \hline
      
{$0.15-0.8 M_\odot$}& {} & {} & {}\\
{All}& {2.8} & {-6.4} & {-1.5} \\
{Compact}& {2.8} & {-5.8} & {-1.5} \\
{Extended}& {2.4} & {-6.6} & {-1.5} \\

{$0.1-0.5 M_\odot$}& {} & {} & {}\\
{All}& {2.1} & {-5.7} & {-1.4} \\
{Compact}& {2.2} & {-5.3} & {-1.4} \\
{Extended}& {1.9} & {-5.7} & {-1.4} \\

{$0.3-0.8 M_\odot$}& {} & {} & {}\\
{All}& {3.8} & {-6.7} & {-1.8} \\
{Compact}& {3.9} & {-6.0} & {-1.8} \\ 
{Extended}& {3.2} & {-6.8} & {-1.8} \\

{$0.5-0.8 M_\odot$}& {} & {} & {}\\
{All}& {4.6} & {-7.2} & {-2.2} \\
{Compact}& {4.6} & {-6.4} & {-2.2} \\ 
{Extended}& {4.0} & {-7.4} & {-2.2} \\

      \hline\hline
    \end{tabular}
  \end{center}
\end{table}

\section{Discussion}\label{discussion}

The near universality in the evolution of $\alpha$ suggests that as a cluster loses mass, the same distribution of stellar masses are either tidally stripped or removed via two-body interactions from the cluster no matter when the escape occurs. This statement is however only applicable to star clusters with dissolution timescales greater than the time scale for stellar evolution, which is true for globular clusters \citep{kruijssen09b}. We do find that the strength of the tidal field and amount of mass segregation a cluster undergoes before becoming tidally affected can influence the stellar mass function evolution, however the effects are minimal. Therefore, it is possible to use the mass function - initial mass relation presented here to determine the initial total stellar mass of any Galactic globular cluster.

\begin{figure}
\centering
\includegraphics[width=\columnwidth]{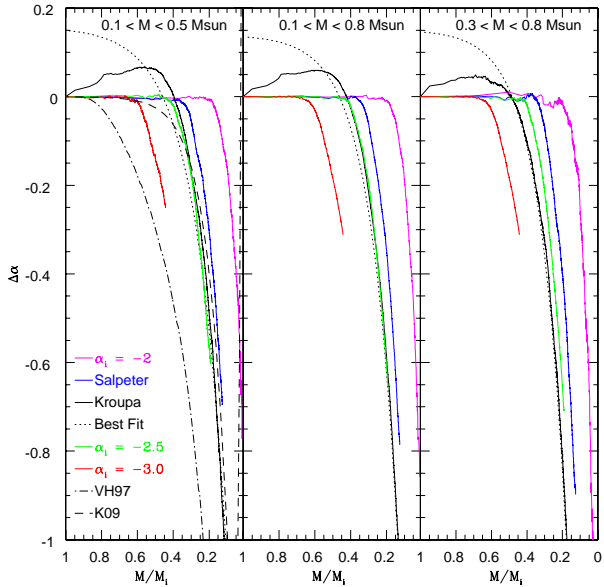}
\caption{Change in the slope of the stellar mass function from its initial value versus the fraction of mass lost for model clusters with different IMFs. Functional forms of the IMF include \citet{kroupa93} (solid black line), \citet{salpeter55} (blue), and power laws with each $\alpha_i$ provided in the legend. At each time step $\alpha$ is calculated using stars within $0.1-0.5 M_\odot$ (left panel), $0.15-0.8 M_\odot$ (centre panel) and $0.3-0.8 M_\odot$ (right panel). In the left panel, the evolution of $\alpha$ as determined by \citet{kruijssen09b} (K09) and \citet{vesperini97} (VH97) are also included.}
  \label{fig:IMF}
\end{figure}

Our definition of initial total stellar mass is best described as the mass that a globular cluster has when its stellar mass function begins to deviate from its initial value. The calculation of the initial mass, since it stems from the simulations presented above, is based on two key assumptions. The first assumption is that all clusters form with the same IMF. While the evolution of the differential mass function has been shown to be independent of the IMF \citep{lamers13}, an IMF would still have to assumed in order to use a cluster's differential mass function to calculate its initial mass. Therefore, to test the dependence of our mass function - initial mass relation on this assumption we present additional simulations of our model cluster with a circular orbit at 6 kpc, initial mass of $6 \times 10^4 M_\odot$, and initial half light radius of 6 pc with a range of IMFs. More specifically we model clusters with power law IMFs that range in $\alpha_i$, with $\alpha_i = -2.35$ equalling a \citet{salpeter55} IMF. To best compare models with different IMFs, we plot the evolution of $\Delta\alpha = \alpha_i - \alpha$ for each cluster in Figure \ref{fig:IMF} with $\alpha$ calculated in three different mass ranges. For the lower mass range, we also compare our results to previous studies by \citet{kruijssen09b} (K09) and \citet{vesperini97} (VH97). 

Figure \ref{fig:IMF} indicates that our mass function - initial mass relation is not independent of the stellar IMF. Clusters with top heavy IMFs are examples of how the evolution of $\alpha$ is different for clusters with short dissolution times \citep{kruijssen09b}, as a significant amount of mass is lost quickly due to stellar evolution while the low mass end of the mass function remains unchanged. The early mass loss also leads to significant cluster expansion and the short cluster dissolution time. Hence $\alpha$ stays relatively constant until the cluster is close to completely evaporating. 

For clusters with a \citet{salpeter55} IMF or power law IMF with $\alpha_i=-2.5$, which are similar to \citet{kroupa93} over the mass range in which $\alpha$ is calculated, Equation \ref{mfrelation} is able to recover the initial mass of a cluster to within $25\%$ and $8\%$ respectively when stars between  $0.15-0.8 M_\odot$ are used. Using the lower mass range provides initial mass estimates comparable to when the full mass range is used, however the higher mass range yields errors approximately $10\%$ higher due to its sensitivity to stellar evolution. If a cluster has a bottom heavy IMF, very little mass loss will occur due to stellar evolution such that $\alpha$ begins to decrease earlier than predicted by Equation \ref{mfrelation}.

Comparison of our mass function - initial mass relation to the works of \citet{kruijssen09b} and \citet{vesperini97} also provides insight into its dependence on the IMF, stellar evolution, and the initial binary fraction. The discrepancy between our models and \citet{vesperini97} is a bit surprising, as their models have a power law IMF with $\alpha_i = -2.5$ and they take into considerate the effects of stellar evolution. Hence the results of \citet{vesperini97} should be directly comparable to our models in Figure \ref{fig:IMF}. The key differences between the two studies are that our maximum stellar mass ($50 M_\odot$) is much higher than \citet{vesperini97} ($15 M_\odot$) and our study also includes an initial binary fraction of $4\%$. A high maximum stellar mass will result in significant early mass loss before $\alpha$ evolves from its initial value. And as previously mentioned, binary stars also serve to delay the evolution of $\alpha$ as binary stars are disrupted and single stars population the low mass end of the mass function. The decrease in $\alpha$ found by \citet{vesperini97} is similar to ours if adjusted for the different fractions of mass lost before $\alpha$ begins to deviate from its initial value. 

Theoretical models for the evolution of $\Delta\alpha$ given by \citet{kruijssen09b} on the other hand are in strong agreement with our own for $\frac{M}{M_i} \le 0.5$ (see Figure \ref{fig:IMF}). The main difference is only in the early evolution of $\Delta\alpha$, which as previously stated is sensitive to the maximum stellar mass and initial binary fraction. Hence both this study and \citet{kruijssen09b} offer methods for calculating the initial mass of a globular cluster based on the change in the slope of the mass function from its initial value, assuming $\frac{M}{M_i} \le 0.5$. Further simulations are required to determine how Equation \ref{mfrelation} depends on the maximum stellar mass and initial binary fraction.

The second assumption associated with our calculation of initial mass stems from the fact that we assume clusters have always had their present day orbit. We have not modelled stellar clusters within their formation environment (where tidal shocks from GMCs have been shown to be a major source of mass loss) or as they migrate from their formation environment to their current orbit. How this early stage of globular cluster evolution affects the evolution of the stellar mass function is still not well understood, and may influence the calculations presented here. However, as discussed in Section \ref{intro} there remain two possibilities in which the effects of early tidal shocks may have a negligible impact on our results. First, if these early tidal shocks remove stars over the entire stellar mass range, then clusters will more or less retain their primordial IMF when they migrate outwards to the halo. Hence the initial mass presented here will simply reflect the cluster's mass once it reaches the halo. If these early tidal shocks acting on globular clusters soon after they form do not remove mass from the cluster in some non-standard way, our calculated initial stellar mass will then effectively represent the cluster's actual initial stellar mass at birth. More specifically, as long as tidal shocks are removing the outermost stars in the cluster the same as tidal evaporation, albeit at an accelerated rate, the evolution in $\alpha$ as a function of the remaining mass fraction should remain the same.  If anything, since early tidal shocks are essentially stronger and more frequently occurring versions of disk shocks, the evolution of $\alpha$ is likely to be faster than presented here. In this case, the initial masses calculated with Equation \ref{mfrelation} are only lower limits. Detailed simulations of clusters in this early phase over a range of initial conditions are required before we can quantify its effect on the mass function - initial mass relation presented here.

\subsection{Application to Milky Way Globular Clusters} \label{mwapp}

\begin{figure}
\centering
\includegraphics[width=\columnwidth]{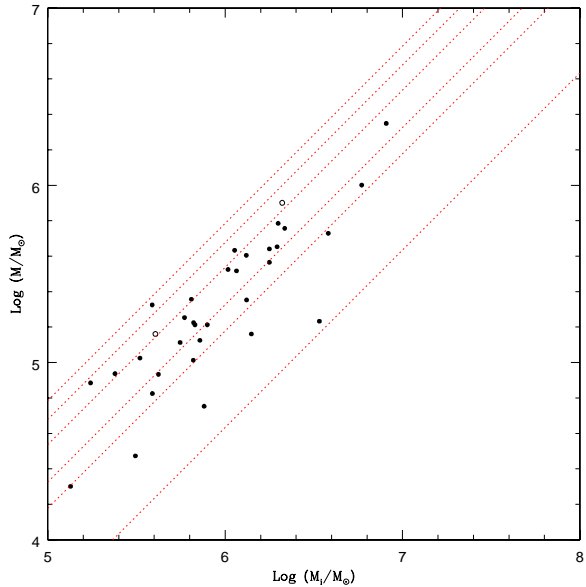}
\caption{Present day globular cluster mass as a function if initial mass calculated via Equation \ref{mfrelation}. Open circles mark clusters for which only an upper limit could be placed on their initial mass. Red dotted lines represent lines of constant $\alpha$ equal to -2.0 (lower red line), -1.5, -1.0, -0.5, 0.0, and 0.5 (upper red line).}
  \label{fig:mw}
\end{figure}

Our study on the evolution of a cluster's stellar mass function can be applied to observations of globular clusters using the observed present-day \textit{global} mass function slope, or even the mass function slope within a narrow radial bin at $r_h$, since this has been shown to be comparable to the global mass function \citep{demarchi07,demarchi10,paust10}. From the datasets of \citet{paust10} and \citet{demarchi10}, combined with the Harris GC catalog \citep{harris96}, we have the present-day total stellar mass (via the integrated V-band magnitude of the clusters) and values of $\alpha$ for 33 Galactic globular clusters. For clusters in both datasets, we take $\alpha$ values from \citet{paust10}. Given Equation \ref{mfrelation} we estimate the initial total stellar mass of each of these clusters and plot their initial and present day total stellar masses in Figure \ref{fig:mw}. Since each $\alpha$ was measured over different mass ranges within $0.1 - 0.8 M_\odot$, we use the most appropriate mass range from Table \ref{table:alphaparam} when applying Equation \ref{mfrelation} to each cluster. The red dotted lines in Figure \ref{fig:mw} mark contours of constant $\alpha$ while open circles represent clusters for which we could only calculate an upper limit for their initial mass. Table \ref{table:mwgc} contains the present day parameters, initial masses and orbital parameters (when possible) of each cluster in our dataset. Orbital parameters are taken from \citet{dinescu99}, \citet{dinescu07} and \citet{dinescu13}. We stress that it is only possible to constrain the total initial \textit{stellar} masses in our sample of globular clusters. Assuming some star formation efficiency $< 100\%$, the total initial cluster masses (including both gas and stars) will have been even higher. 

\begin{table}
  \caption{Initial Masses of Galactic Globular Clusters}
  \label{table:mwgc}
  \begin{center}
    \begin{tabular}{lcccccc}
      \hline\hline
      {$ID$} &{$M$} & {$\alpha$} & {$M_i$} & {Rp} & {Ra} \\
      {} &{($log\frac{M}{M_\odot}$)} & {} & {($log\frac{M}{M_\odot})$} & {(kpc)} & {(kpc)} \\

      \hline

 Pal 5 & 4.30 & -0.4 & 5.13 & {} & {} \\
 NGC 104 & 6.00 & -0.84  & 6.77  & 5.2  & 7.3 \\
 NGC 288 & 4.93 & -0.83  & 5.62  & 1.7 & 11.2 \\
 NGC 362 & 5.60 & -1.69  & 6.12 & 0.8 & 10.6 \\
 NGC 1261 & 5.35 & -0.59 & 6.12  & {} & {} \\
 NGC 1851 & 5.56 & -0.85 & 6.23 & 5.7 & 30.4 \\
 NGC 2298 & 4.75 & 0.5 & 5.88 & 1.9 & 15.3\\
 NGC 3201 & 5.21 & -0.77 &  5.90  & 9.0 & 22.1 \\
 NGC 5053 & 4.94 & -1.46 &  5.38  & {} & {} \\
 NGC 5139 & 6.35 & -1.2 & 6.91 & 1.2 & 6.2 \\
 NGC 5272 & 5.78 & -1.31 &  6.30  & 5.5 & 13.4 \\
 NGC 5286 & 5.73 & -0.32 &  6.58  & {} & {} \\
 NGC 5466 & 5.02 & -1.15 &  5.52  & 6.6 & 57.1 \\
 NGC 5904 & 5.76 & -1.15 &  6.33  & 2.5 & 35.4 \\
 NGC 5927 & 5.36 & -1.44 &  5.81  & 4.5 & 5.5 \\
 NGC 6093 & 5.52 & -1.36 &  6.02  & 0.6 & 3.5 \\
 NGC 6121 & 5.11 & -1.00 & 5.75 & 0.6 & 5.9 \\
 NGC 6205 & 5.65 & -0.98  &  6.29  & 5.0 & 21.5 \\
 NGC 6218 & 5.16 & 0.1 & 6.15 & 2.6 & 5.3 \\
 NGC 6254 & 5.22 & -1.1 & 5.82 & 3.4 & 4.9\\
 NGC 6304 & 5.16 & -1.85 & 5.61 & {} & {}  \\
 NGC 6341 & 5.52 & -1.23  &  6.06  & 1.4 & 9.9 \\
 NGC 6352 & 4.82 & -0.6 & 5.59 & {} & {} \\
 NGC 6362 & 5.01 & -0.49  &  5.82  & 2.4 & 5.5 \\
 NGC 6397 & 4.88 & -1.6 & 5.24 & 2.34 & 6.0 \\
 NGC 6496 & 5.12 & -0.7 & 5.86 & {} & {}  \\
 NGC 6541 & 5.64 & -1.07  &  6.25  & {} & {} \\
 NGC 6656 & 5.63 & -1.5 & 6.05 & 2.76 & 8.76 \\
 NGC 6712 & 5.23 & 0.9 & 6.53 & 0.9 & 6.2 \\
 NGC 6752 & 5.32 & -1.7 & 5.59 & {} & {}  \\
 NGC 6809 & 5.25 & -1.3 & 5.77 & 1.9 & 5.8 \\
 NGC 6838 & 4.47 & 0.2 & 5.49 & 4.8 & 6.7 \\
 NGC 7078 & 5.90 & -1.9 & 6.32 & 5.4 & 10.3 \\
 NGC 7099 & 5.21 &  -0.92  &  5.83  & 3.0 & 6.9 \\
 
       \hline\hline
    \end{tabular}
  \end{center}
\end{table}

It should be noted that only upper limits could be placed on the initial masses of NGC 6304 and NGC 7078 because their present day $\alpha$ is less than $\alpha_i$ in the $0.3-0.8 M_\odot$ mass range. The simplest explanation for this discrepancy is that the $0.3-0.8 M_\odot$ mass range encompasses the bend in the \citet{kroupa93} IMF at $0.5 M_\odot$, which can strongly affect measurements of $\alpha$ depending on the cluster's dynamical age. To place an upper limit on each cluster's initial mass, we use Equation \ref{mfrelation} with the $0.5-0.8 M_\odot$ mass range since $\alpha_i$ is less than each cluster's present day $\alpha$. Followup measurements of $\alpha$ between $0.1-0.5 M_\odot$ or $0.5-0.8 M_\odot$ will allow for a proper estimation of the initial masses of NGC 6304 and NGC 7078.

Figure \ref{fig:mw} suggests that the clusters in our data set were on average a factor of 4.5 times more massive than they are today, with some clusters having initial masses greater than a factor of 10 times (e.g. NGC 2298 and NGC 6838) or even 20 times (e.g. NGC 6712) larger than their present day values. Based on our limited data set, we estimate that the initial globular cluster mass function was centred around $LOG(\frac{M}{M_\odot}) = 6.0$. However, it should be stressed that this is the initial globular cluster mass function of clusters that survived to the present day only, and does not include clusters that have dissolved due to relaxation, dynamical friction, or disk and bulge shocks \citep{gnedin99}.  The inclusion of these clusters will change the initial globular cluster mass function substantially \citep{kruijssen09c}, and should ultimately contribute to lowering our calculated mean mass.

Given the mass function of a globular cluster, our parameterization (Equation \ref{mfrelation}) provides an estimate of initial cluster mass as a function of its present day mass function slope. However, the implied mass loss is on average a factor of 2 larger than theoretical predictions for the dynamical mass loss experienced by Galactic clusters on their observed present-day orbits \citep{kruijssen09a, rossi15}. Hence there is a discrepancy between our empirical calculation of the fraction of mass lost by a globular cluster using the present-day mass function slope compared to the predicted fraction of mass lost from N-body simulations (for a given orbit). However it should be noted that we estimate initial masses less than \citet{kruijssen09a} for NGC 288, NGC 5139, NGC 6093, NGC 6121, NGC 6809, and NGC 7089. The latter two clusters having initial masses within $10 \%$ of the estimated \citet{kruijssen09a} values. This discrepancy, and the possible explanations behind it, are likely either due to real Galactic clusters not being subject to a comparable tidal field as our $N$-body simulations, or due to having initial parameters outside the parameter space covered in this study (e.g. small initial cluster sizes or short perigalactic distances ). As discussed in \citet{marks08}, primordial gas expulsion may also help to unbind a significant fraction of low-mass stars in initially mass segregated clusters very early on in their evolution, and this effect is not accounted for in our simulations.

We know that our model clusters have not experienced the same tidal field as actual globular clusters since A) we have not modelled clusters evolving in their early formation environment and B) many clusters have orbits outside of our parameter space \citep{dinescu99, dinescu07, dinescu13}. Including the additional mass loss mechanisms associated with a cluster's early formation environment, specifically tidal shocks from encounters with GMCs and the unbinding of stars via gas expulsion, could explain why these clusters appear to have lost a significant amount of mass. However, we caution that the evolution of $\alpha$ prior to a cluster's migration to the halo has not been explored, and may increase the uncertainty of the mass function - initial mass relation presented here.

Modelling additional clusters with very small values of $R_p$ may partially explain the high initial masses of these clusters, as they are subject to an extremely strong tidal field and higher mass loss rates than the clusters modelled in the present study. However \citet{kruijssen09a} estimates that the mass loss rate due to tidal evaporation and disk shocking associated with clusters with small $R_p$ values still yield initial masses that are less than presented here. Similar results were obtained by \citet{baumgardt03} and \cite{baumgardt08}. While a more detailed treatment of the Galactic potential at small $R_{gc}$ may partially decrease the discrepancy, it seems that the additional mass loss mechanisms discussed above will still be necessary to explain the high initial cluster masses associated with the empirically-measured stellar mass functions.

Finally, it is possible that many observed clusters formed extremely compact compared to any of our model clusters, or were just recently placed on their current orbits. An extremely compact cluster will have time to almost completely mass segregate before becoming tidally filling and would therefore have a higher present day $\alpha$ than our suite of simulations would predict. Similarly, if these clusters spent the majority of their lifetimes in a weak tidal field and were able to expand and relax before orbiting within the inner regions of the Milky Way, we would also observe a higher than expected $\alpha$.

\section{Summary} \label{summary}

We use a large suite of $N$-body simulations of star cluster evolution to determine the effects of initial mass, initial size and orbit on the evolution of a cluster's stellar mass function. We illustrate that the observed slope of a cluster's present day stellar mass function $\alpha$ is primarily dependent on the fraction of stellar mass that a cluster has lost and is approximately independent of its initial mass, initial size, orbital distance, eccentricity, and inclination. Our work confirms the results of previous numerical and analytical studies on the evolution of star clusters with long disruption time-scales \citep{vesperini97, kruijssen09b}. We note that there is some scatter about this relation due to the cluster orbit, as clusters that are subject to stronger tidal fields, tidal heating or tidal shocks can be more easily stripped of high mass stars, as found by \citet{webb14a}. There is also a secondary dependence of the stellar mass function's evolution on the initial size of a cluster, since initially compact clusters have a chance to relax and mass segregate as they expand to fill their tidal radius. Due to the cluster undergoing some degree of mass segregation before it starts to be tidally stripped, more low mass stars will be removed from the cluster and the slope of the mass function will increase faster. It also appears that orbital inclination can alter the evolution of $\alpha$, as increased tidal heating due to passages through the Galactic disk can remove a larger fraction of higher mass stars from a cluster than if it orbited in the plane of the disk.

Since the scatter induced by these factors is minimal, our models can be used to establish an analytic relationship between $\alpha$ and the fraction of mass that a cluster has lost (Equation \ref{mfrelation}) that is accurate to within $\sim 7\%$. If the initial size of a cluster can be constrained via other means, the uncertainties on the initial masses can be decreased to $\lesssim 5\%$. However, we note that our mass function - initial mass relation is based on the assumptions that all clusters form with a Universal \citet{kroupa93} IMF and that forms of mass loss not included in our models, mainly tidal shocks experienced by clusters in their formation environment, do not significantly alter the evolution of $\alpha$. Additional simulations find that as long as the high-mass end of the IMF has $ -2.6 \le \alpha_i \le -2.3$, our fitting functions are accurate. Clusters with top heavy IMFs tend to dissolve quickly with the stellar mass function undergoing very little change for stars between $0.15-0.8 M_\odot$. Clusters with bottom heavy IMFs lose very little mass due to stellar evolution such that $\alpha$ begins increasing much earlier than our mass function - initial mass relation predicts. Comparisons to previous studies also suggest the maximum stellar mass and initial binary fraction may also influence the evolution of $\alpha$. Future simulations are necessary to accurately model how these factors, as well tidal shocks experienced by clusters in their formation environment, will affect the early evolution of $\alpha$.

We use our mass function - initial mass relationship to determine the initial total stellar mass of any Galactic globular cluster that has a known mass function and present day mass. Combining our mass function - initial mass relation with the 33 Galactic globular cluster dataset of \citet{paust10}, we find surviving globular clusters were on average a factor of 4.5 times more massive than they are seen today, with some clusters being greater than a factor of 10 times more massive. We note that for two clusters our relation could only be used to put upper limits on their initial masses. We caution that these initial mass estimates are higher than other predictions in the literature based on mass loss due to tidal evaporation and disk shocking alone. This inconsistency is also apparent from the $N$-body simulations we use to calibrate the relation between $\alpha$ and $M/M_{i}$(Equation \ref{mfrelation}). If we extrapolate these simulations to globular cluster masses, they yield a total dynamical mass loss smaller than Equation \ref{mfrelation} would predict for the observed mass function slopes of Galactic globular clusters. However, when scaled to globular cluster masses, our models would fail to reproduce the observed distribution of stellar mass functions as they will have only lost a small percentage of their initial mass. This discrepancy suggests that additional mass loss mechanisms, like tidal shocks while in the early formation environment, are necessary to explain the present day mass functions of some Galactic globular clusters. Our predicted initial masses would then only be in disagreement with previous studies if these additional mass loss mechanisms do not alter the evolution of the mass function such that our predictions correspond to the initial mass of clusters once they have migrated into the halo. However if these additional mass loss mechanisms affect the evolution of a cluster's mass function the same way that mass loss via tidally-limited two-body relaxation does, then our predicted initial masses actually correspond to the true initial mass of each cluster at formation. The possibility still remains, however, that additional mass loss mechanisms may alter the evolution of the mass function in some non-standard way, in which case the connection between our predicted initial masses and a cluster's true initial mass becomes unclear. The effects of additional mass loss mechanisms on the evolution of $\alpha$ need to be explored in order to determine how they influence the mass function - initial mass relation.

Having a simple method to calculate initial cluster masses that is based on two observable parameters will help future studies to further constrain the GCIMF and initial distribution of cluster sizes. These are two very important parameters when studying both globular cluster and galaxy formation models. Our method provides a simple prescription for the evolution of $\alpha$ that can be adopted by fast codes wishing to model the evolution of large cluster populations \citep[e.g.]{alexander12, alexander13, alexander14} or be used to suggest initial parameters for simulations attempting to model individual globular clusters. 

\section{Acknowledgements}
JW and NL would like to thank our referee, Diederik Kruijssen, for his detailed and constructive comments which served to improve our study. JW and NL acknowledge financial support through research grants and scholarships from the Natural Sciences and Engineering Research Council of Canada. The authors would also like to thank Alison Sills, Bill Harris, and Mark Gieles for helpful discussions and comments on the manuscript. This work was made possible by the facilities of the Shared Hierarchical Academic Research Computing Network (SHARCNET:www.sharcnet.ca) and Compute/Calcul Canada. The authors are also grateful to Dr. Emmett Brown for useful discussions.


\bsp

\label{lastpage}

\end{document}